\documentclass[prb,aps,10pt,twocolumn,amsmath,amssymb,superscriptaddress,floatfix]{revtex4-2}   
\usepackage{lineno}
\usepackage{graphicx}
\usepackage{dcolumn}
\usepackage{bm}
\usepackage{rotating}
\usepackage{subfigure}
\usepackage{pstricks}
\usepackage{braket}
\usepackage{pst-node}
\usepackage{overpic}
\usepackage{amsmath}
\usepackage{color}
\usepackage{makecell}
\usepackage{booktabs}
\usepackage{flushend}
\usepackage{balance}
\usepackage{upgreek}
\usepackage{enumerate}
\usepackage[colorlinks,
			linkcolor=blue, 
			urlcolor=blue, 
			anchorcolor=blue, 
			citecolor=blue]{hyperref} 

\begin{document}

\preprint{APS/123-QED}

\title{The nanoscale imaging of the bulk polycrystalline material with the effects of depth of field and field of view based on x-ray free electron laser}

\author{Chuan Wang}
\affiliation{College of Physics, Sichuan University, Chengdu 610064, People's Republic of China}
\affiliation{Key Laboratory of Radiation Physics and Technology, Ministry of Education, Chengdu 610064, People's Republic of China}
\affiliation{Key Laboratory of High Energy Density Physics and Technology, Ministry of Education, Chengdu 610064, People's Republic of China}

\author{Yihan Liang}
\affiliation{National Key Laboratory of Shock Wave and Detonation Physics, Institute of Fluid Physics, China Academy of Engineering Physics, Mianyang 621900, People's Republic of China}

\author{Ronghao Hu}
\affiliation{College of Physics, Sichuan University, Chengdu 610064, People's Republic of China}
\affiliation{Key Laboratory of Radiation Physics and Technology, Ministry of Education, Chengdu 610064, People's Republic of China}
\affiliation{Key Laboratory of High Energy Density Physics and Technology, Ministry of Education, Chengdu 610064, People's Republic of China}

\author{Kai He}
\author{Guilong Gao}
\author{Xin Yan}
\author{Dong Yao}
\author{Tao Wang}
\affiliation{Xi’an Institute of Optics and Precision Mechanics (XIOPM), Chinese Academy of Sciences (CAS), Xi’an 710119, People's Republic of China}

\author{Xiaoya Li}
\affiliation{National Key Laboratory of Shock Wave and Detonation Physics, Institute of Fluid Physics, China Academy of Engineering Physics, Mianyang 621900, People's Republic of China}

\author{Jinshou Tian}
\affiliation{Xi’an Institute of Optics and Precision Mechanics (XIOPM), Chinese Academy of Sciences (CAS), Xi’an 710119, People's Republic of China}

\author{Wenjun Zhu}
\affiliation{National Key Laboratory of Shock Wave and Detonation Physics, Institute of Fluid Physics, China Academy of Engineering Physics, Mianyang 621900, People's Republic of China}

\author{Meng Lv}
\email[Corresponding author: ]{lvmengphys@scu.edu.cn}
\affiliation{College of Physics, Sichuan University, Chengdu 610064, People's Republic of China}
\affiliation{Key Laboratory of Radiation Physics and Technology, Ministry of Education, Chengdu 610064, People's Republic of China}
\affiliation{Key Laboratory of High Energy Density Physics and Technology, Ministry of Education, Chengdu 610064, People's Republic of China}

\begin{abstract}
	We study the effects of the depth of field (DoF) and field of view (FoV) of the optical lens and extract the scattered light of the region to be imaged within the bulk polycrystalline material based on the objective BCDI. We describe how the DoF and FoV quantitatively limit the diffraction volume, where the DoF and FoV limit the scattering region parallel and perpendicular to the direction of the light source respectively. We demonstrate this scheme by simulating the separate imaging of a submicron-sized crack region within a few $\mathrm{\upmu m}$-sized Si bulk material, and obtain a high imaging quality. This scheme enables imaging of selected regions within bulk polycrystalline materials with the resolution up to the order of 10 nm.
\end{abstract}

\maketitle

\section{Introduction}
Materials subjected to dynamic compression can induce strongly nonlinear processes on characteristic submicron spatial scales, such as defect activation, twin formation, and phase transitions. These processes are combined with other factors in subsequent evolution, including shock wave loading and unloading processes, which largely determine the material's subsequent behavior \cite{Kraus2017,doi:10.1126/science.1135080, 2015Schropp, BURGERS1946, PhysRevLett.67.3412}. Microscopic imaging of materials is crucial for understanding their response to dynamic compression. However, the extreme conditions of the loading method often limit the sample size to a lower limit of 100 $\mathrm{\upmu m}$, which is usually the upper limit of micro-scale for most solid materials. Therefore, it is necessary to develop direct imaging methods for the formation and motion of submicron scale interfaces within tens to hundreds of $\mathrm{\upmu m}$-sized bulk polycrystalline materials using three-dimensional (3D) microscopy and tomography.

Small angle x-ray scattering provides statistical information about the structure of materials on a nanometer to micron scale in the region where x-rays interact with the material \cite{2009Hura, RENAUD2009255}. Dynamic x-ray diffraction measures changes in the crystal structure of materials under dynamic compression, providing insights into deformation and phase transformation \cite{doi:10.1080/10408347.2014.949616, doi:https://doi.org/10.2136/sssabookser5.1.2ed.c12}. X-ray imaging techniques based on brightness detection (e.g., x-ray absorption imaging) and phase detection (e.g., x-ray phase contrast imaging) offer high spatial and density resolution, such as XFEL-based phase contrast imaging with submicron spatial resolution, and can investigate processes such as material density distribution and interfacial instability under dynamic compression \cite{TIDWELL2000285, Burvall:11, 10.1017/S1431927615011617, doi:10.1126/sciadv.aau8044, Hagemann:il5056}. However, none of these methods yield quantitative structural information. Coherent diffraction imaging (CDI) is a powerful imaging technology that has opened new research directions in physics, material science, and bioscience \cite{Fienup:82, 1999Miao, ALLEN200185, PhysRevLett.89.088303, Bauschke:02, PhysRevLett.93.023903, BELEGGIA200437, PhysRevLett.97.215503, Guizar-Sicairos:08, MAIDEN20091256, doi:10.1126/science.aaa1394, 2009Harder,  Latychevskaia:12, Zurch:13, Pein:16, Hirose:17, PhysRevLett.90.175501, doi:10.1073/pnas.0503305102, 2010Nelson, 2010jiang, PhysRevLett.87.195505, Coughlan:gm5048, PhysRevApplied.14.024085, PhysRevB.103.214103, PhysRevB.103.014102, PhysRevLett.98.034801, Wittwer:22, BAROLAK2022113418, Yu2022, PhysRevLett.97.025506, Guizar-Sicairos:11, Li2021, 2011Roy, 2012Sun}. Bragg coherent diffraction imaging (BCDI) is a non-destructive imaging technique that can study dislocation dynamics, stacking faults, and twinning at the submicron scale by imaging the shape and displacement fields of 3D crystals from Bragg peaks \cite{doi:10.1126/science.aam6168, Ulvestad2017, Favre_Nicolin_2010, Yang2013, doi:10.1063/5.0005771, PhysRevLett.117.225501, Hruszkewycz:15, PhysRevB.93.184105, PhysRevA.99.053838}. However, classical BCDI is only suitable for submicron-sized isolated crystals \cite{Pedersen:18, PhysRevResearch.2.033031}. Bragg ptychography can image extended crystals, but its limited longitudinal coherence restricts the thickness of crystals to no more than 1 $\mathrm{\upmu m}$, and it is not compatible with dynamic imaging \cite{PhysRevLett.98.034801, Mastropietro2017}. Near-field CDI provides spot sizes ranging from tens to hundreds of $\mathrm{\upmu m}$, allowing the study of hundreds of $\mathrm{\upmu m}$-sized samples with a resolution of hundreds of nm, but this resolution is insufficient for studying structural information at the submicron scale \cite{PhysRevB.89.184105, PhysRevB.96.054104}. Although the lensless method is a significant advantage of classical CDI, it is worth revisiting lens-based CDI. In the reporter-based imaging scheme proposed by Soltau \emph{et al.} \cite{PhysRevLett.128.223901}, multilayer zone plates are used as an objective lens to achieve better resolution than classical CDI signals. Pedersen \emph{et al.} \cite{Pedersen:18, PhysRevResearch.2.033031} propose the objective BCDI and demonstrate that the introduction of a set of lenses does not change the intensity distribution in the far field. However, the analysis of how to control the size and location of the imaging region is not mentioned in their work. The ability to control the size and location of the imaged region within the bulk polycrystalline material is important for applications such as imaging shockwave fronts after shock wave action or imaging material responses under dynamic compression.  Additionally, it is difficult to directly filter out x-rays scattered from the sample outside the field of view (FoV) using lenses for submicron scale imaging within tens to hundreds of $\mathrm{\upmu m}$-sized bulk materials. Villanueva-perez \emph{et al.} \cite{Villanueva-Perez:18} propose the x-ray multi-projection imaging (XMPI) scheme, which uses a beam splitter to generates a number of beams by Laue diffraction to illuminate samples from different angles at the same time.
  
In this paper, aiming at the demand for submicron scale imaging in tens to hundreds of $\mathrm{\upmu m}$-sized bulk materials, we firstly study the effects of depth of field (DoF) and FoV of the optical lens to extract the scattered light of the desired imaging region within the bulk polycrystalline material. The size and position of the imaging region are selected by designing lens parameters and geometric configuration. Then, we demonstrate the 3D imaging of a submicron-sized crack region within a several $\mathrm{\upmu m}$-sized Si bulk material. Finally, we discuss the possibility of extending this method to 3D single-shot imaging of mesoscale materials based on x-ray free electron laser (XFEL).

\section{Method}

\begin{figure*}[htbp]
	\centering
	\includegraphics[width=0.9\textwidth]{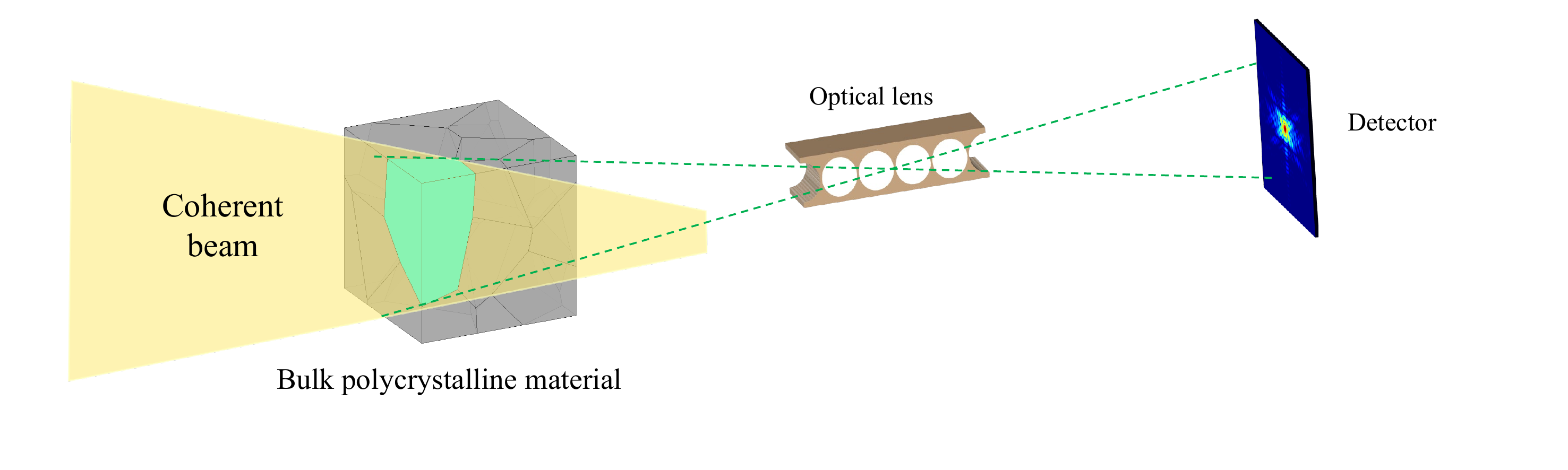}
	\caption{\label{fig1}The schematic of objective BCDI geometry, which describes the use of optical lens to extract the scattered light of the selected region in the bulk polycrystalline material for imaging.}
\end{figure*}

In the BCDI scheme, a nanocrystal is illuminated with a coherent x-ray beam of wavelength $\lambda$ and the diffraction pattern surrounding a Bragg peak is measured in the far field. The incident beam wave vector $\mathbf{k}_{\mathbf{i}}$, exit beam wave vector $\mathbf{k}_{\mathbf{f}}$ ($|\mathbf{k}_{\mathbf{i,f}}|=2\pi /\lambda$), and the nanocrystal is oriented such that the scattering vector $\mathbf{q}=\mathbf{k}_{\mathbf{f}}-\mathbf{k}_{\mathbf{i}}$ is in the vicinity of a Bragg reflection at the reciprocal lattice point $\mathbf{G}_{\mathbf{hkl}}$ \cite{PhysRevLett.117.225501}. The spatial coherence requirement of CDI means that the thickness limit of object is usually about 1 $\mathrm{\upmu m}$. Therefore, for bulk polycrystalline materials, crystals outside the illumination region of the coherent beam will produce diffraction patterns independently, resulting in incoherent superposition with the diffraction pattern of the crystal in the illumination region, i.e., the superposition of intensity rather than the superposition of complex amplitude. Even if all the diffraction crystals are contained in the coherence volume, it can be expected that the coherent superposition of the diffraction intensity of all crystals will produce highly distorted diffraction pattern, for which the support conditions are difficult to play a role in the reconstruction process \cite{Pedersen:18}. The introduction of lens may be a feasible solution to the imaging of bulk polycrystalline materials. As shown in Fig. \hyperref[{fig1}]{1}, objective BCDI uses optical lens to extract scattered light of selected region in bulk polycrystalline material for imaging. By designing the parameters of the lens and the diffraction geometry, we can use the properties of the DoF and FoV of the lens to reduce the effective diffraction volume, in which the DoF limits the diffraction region parallel to the direction of the coherent beam, and the FoV limits the diffraction region normal to the direction of the coherent beam. 

\begin{figure*}[htbp]
	\centering
	\includegraphics[width=0.9\textwidth]{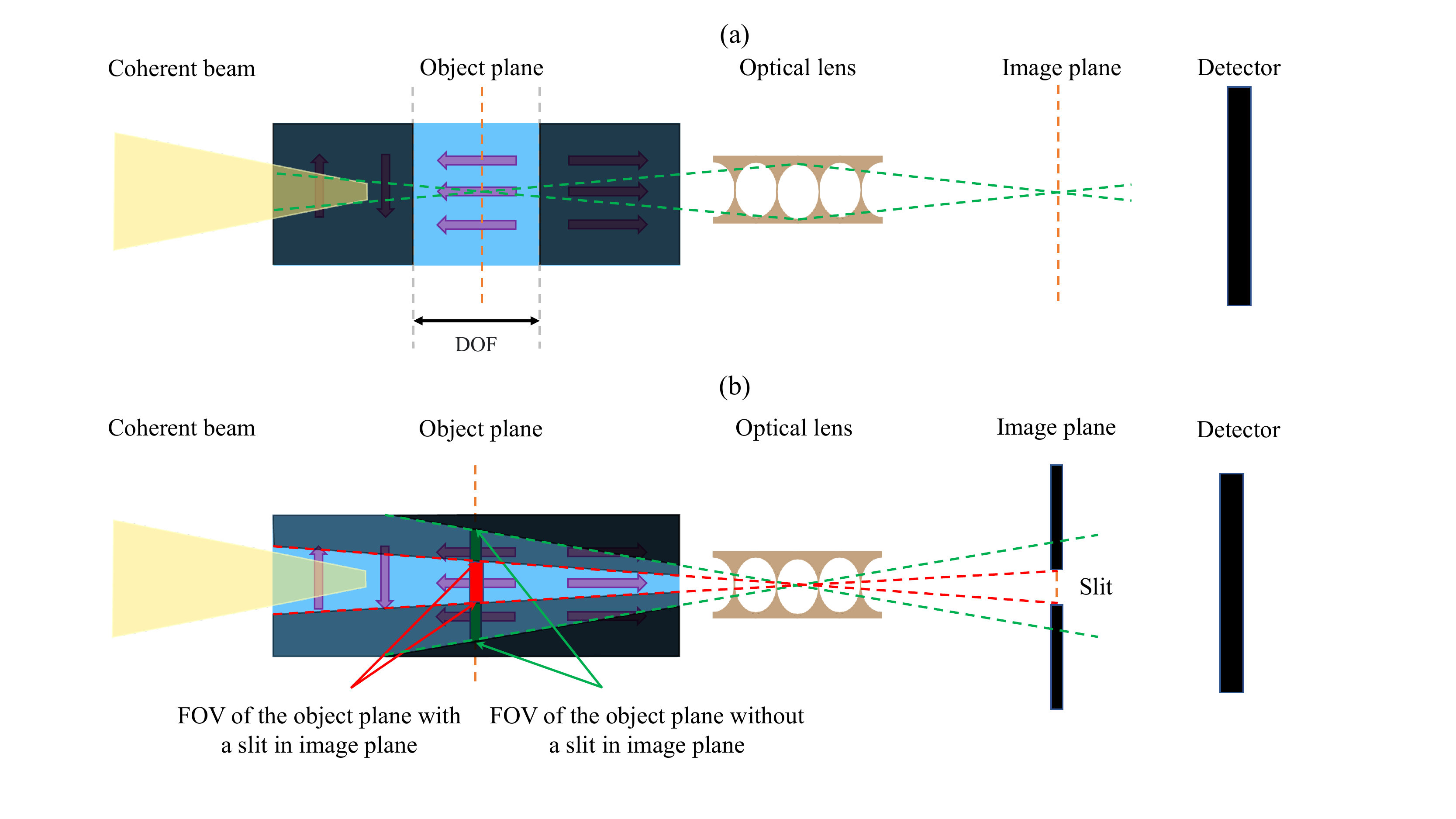}
	\caption{\label{fig2}The schematic diagram of the diffraction volume limited by the DoF and FoV of the lens. (a) The DoF limits the diffraction region parallel to the direction of the coherent beam, and (b) the FoV limits the diffraction region normal to the direction of the coherent beam. In the real image configuration, a slit can be inserted in the image plane to further limit the FoV.}
\end{figure*}

The principle is shown schematically in Fig. \hyperref[{fig2}]{2}. For the convenience of narration, we simplify the object into 2D in the schematic diagram. As shown in Fig. \hyperref[{fig2}]{2(a)}, we set four kinds of identifying cracks along the direction of the coherent beam (the arrows along the four directions of the upper, lower, left and right, respectively, with the size of several hundred nm). By setting the DoF of the lens close to the size of the crack, we can limit the diffraction region along the direction of the coherent beam to the size of the crack region. The position of the object plane can be changed by changing the diffraction geometry (the case demonstrated here is that the restricted diffraction region is the crack region along the left direction). In the direction normal to the coherent beam, the region within the green dotted line is the FoV without slit, and the FoV of the object side is the region marked by the green thick solid line, as shown in Fig. \hyperref[{fig2}]{2(b)}. At this time, the crack regions of the three arrows along the left are all in the FoV. If one want to limit the diffraction region to the crack region in the middle, we can reduce the FoV by changing the geometric configuration. For real image configuration, another method is to add a slit in the image plane. The limited FoV is shown by the region within the red dotted line, and the corresponding FoV of the object plane is the region marked by the green thick solid line. Only the cracked region of the middle left arrow is in the FoV. In this case, we achieve the separate imaging of selected regions within a bulk polycrystalline material.

The imaging of thick sample requires the higher energy x-rays, which means that compound refractive lens (CRLs, the linear arrays of $N$ refractive lenslets) is the optics of choice, and the focal length of CRL can be changed by adjusting the number of lens, which facilitates the optimization of CRL geometry to analyze numerical aperture (NA) and spatial resolution, and others \cite{Pantell:03}. However, thin lens approximation is not applicable to CRL. Based on the ray-transfer matrices, Simons \emph{et al.} provide the closed analytical expressions for critical imaging parameters such as NA, spatial acceptance (vignetting) for both thin- and thick-lens imaging geometries \cite{Simons:ie5162}. The focal length $f_N$ of the CRL of $N$ identical small lens with apex radius of curvature $R$, aperture $2Y$ and center-to-center distance between successive lenslets $T$ (thus $Y=\sqrt{RT}$) is:
\begin{equation}
	f_N = f\varphi \cot\left(N\varphi\right),
\end{equation}
where $f=R/(2\delta)$ (where $\delta$ is the refractive decrement) is the focal length of each lens and the parameter $\varphi=\sqrt{T/f}$. And the imaging condition is:
\begin{equation}
	\frac{1}{d_{1}}+\frac{1}{d_{2}}-\frac{1}{f_{N}}+\frac{f \varphi \tan (N \varphi)}{d_{1} d_{2}}=0,
\end{equation}
where $d_1$ is the object distance and $d_2$ is the image distance. Then the magnification $\mathcal{M}$ of the imaging system can be expressed as:
\begin{equation}
	\mathcal{M} = \frac{d_2 \sin(N\varphi)}{f\varphi}-\cos(N\varphi).
\end{equation}
$d_1$ is a free parameter that determines whether we operate using a real ($d_1>f_N$) or virtual ($d_1<f_N$) image configuration. At the center of the FoV, the NA can be expressed as:
\begin{equation}
	\rm{NA} = 2.35 \sigma_a,
\end{equation}
where $\sigma_a$ is the root mean square width of the Gaussian angular acceptance function, which restricts the spatial resolution $r$ of the final imaging by:
\begin{equation}
	r = \frac{0.61\lambda}{\rm{NA}}.
	\label{resolution}
\end{equation} 
The expression of DoF $\Delta$ and FoV $\Lambda$ can be obtained through simple geometric operation:
\begin{equation}
	\begin{gathered}
		\Delta = \frac{4Yf_Nd_1^2z}{4Y^2f_N^2-d_1^2z^2},\\
		\mathrm{\Lambda_o} = \frac{d_1}{d_2} \mathrm{\Lambda_i},
	\end{gathered}
\end{equation}
where $z$ is the size of the permissible circle of confusion, and $\mathrm{\Lambda_o}$ and $\mathrm{\Lambda_i}$ are the FoV of object and image plane respectively.

The objective BCDI adds CRL as an objective lens between the sample and the detector in the classical BCDI geometry, so the effects of CRL should be considered in the propagation mode of scattered wave field. The fractional Fourier transform (FrFT) \cite{Pedersen:mo5174} has no sampling requirements and it is fast, making it suit for propagation along finite distances or through CRLs. For a given wave field in any plane, it can compute 2D wave fields propagating to any other plane. In our simulations, a series of 2D exit fields are generated along the rocking curve, each of which is multiplied by the vignetting function of the sample plane and propagated through a FrFT to the exit plane of the CRL, where the pupil function is multiplied and the cumulative phase error of the CRL is taken into account by using a phase distortion model characterized by a radially symmetric cosine function with an amplitude of $2\pi$ and a period of 0.5 mm. Finally, the field propagates to the detector plane through the second FrFT. The effects of CRL is characterized by optical transfer function and phase difference in this propagation model and the detail information can be found in Refs. \cite{Simons:ie5162, Pedersen:18}. The finite bandwidth of x-ray leads to its finite coherence length, as well as the effect of chromatic aberration of the optical lens. As a technique applied to XFEL, and to simplify the model, we assume that the x-ray is monochromatic and fully coherent, and that the coherent beam has a sufficiently high brightness and coherent flux, i.e., a sufficiently high signal-to-noise ratio, ignoring the effect of noise on the resolution.

The combination of error reduction (ER) and hybrid input-output (HIO) algorithm outperform either of the algorithms separately in BCDI reconstruction, and is particularly effective in avoiding stagnation and achieving rapid convergence \cite{Fienup:82, doi:10.1063/1.2403783, doi:10.1063/1.5054294}. This hybrid algorithm includes many iteration cycles, one of which is usually composed of dozens of iterations of HIO algorithm, and then 5 to 10 iterations of ER algorithm. Support is also an important factor in the phase retrieval, as it will significantly affect how the electron density of the image is modified in each iteration. A loose support may lead to non-unique solutions to the reconstruction \cite{Fienup:87}. The iterative algorithms iterate between real and reciprocal space, applying various a \emph{priori} constraints in each domain, yet such \emph{priori} knowledge is not always available. Marchesini \emph{et al.} \cite{PhysRevB.68.140101} developed an method called shrinkwrap, which allows for the support size to be dynamically determined as the iterations of the algorithm proceed, thus eliminating the need for \emph{priori} knowledge for reconstruction. In a shrinkwrap reconstruction, the first support can be defined from the auto-correlation function of the object, which can be obtained by the IFT of the diffraction pattern. Although both the correct object density and its centro-symmetry inversion fit within this initial support, the inversion symmetry is progressively lost as the algorithm converges. After a certain number of iterations of the algorithm (HIO/ER), the current reconstructed object is convolved with a Gaussian of width $\sigma_{\mathrm{max}}$ (full width at half maximum of $2.35\sigma_{\mathrm{max}}$) to find the new support. The next round of iterations can then be launched with the new support. This process continues, and the Gaussian width is set to gradually reduce (usually reduced by $1\%$ per update) to support updates until either a stable solution is finally achieved or the specified minimum width $\sigma_{\mathrm{min}}$ is reached. 

\section{Results}

\begin{figure*}[htbp]
	\centering
	\includegraphics[width=0.9\textwidth]{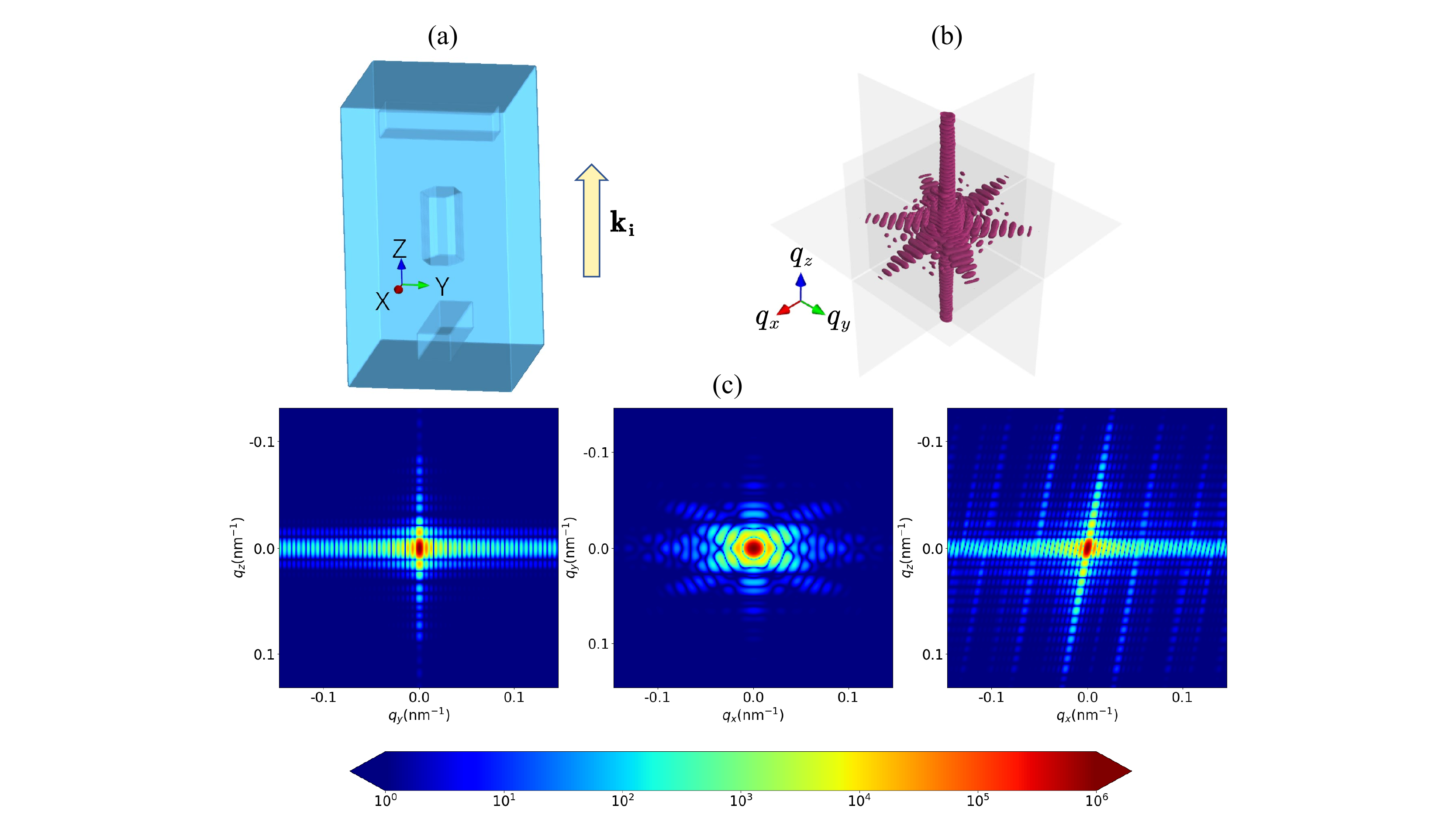}
	\caption{\label{fig3}(a) The density isosurface of the bulk polycrystalline Si material of size $2\mathrm{\upmu m}\times 2\mathrm{\upmu m}\times 4\mathrm{\upmu m}$. (b) The 3D diffraction intensity obtained by stacking a series of 2D slices near the (111) Bragg peak obtained along the rocking curve. (c) The three central slices of the simulated 3D intensity. The clear symmetry of the diffraction intensity indicates that the crystal is strain-free.}
\end{figure*}

\begin{figure*}[htbp]
	\centering
	\includegraphics[width=0.9\textwidth]{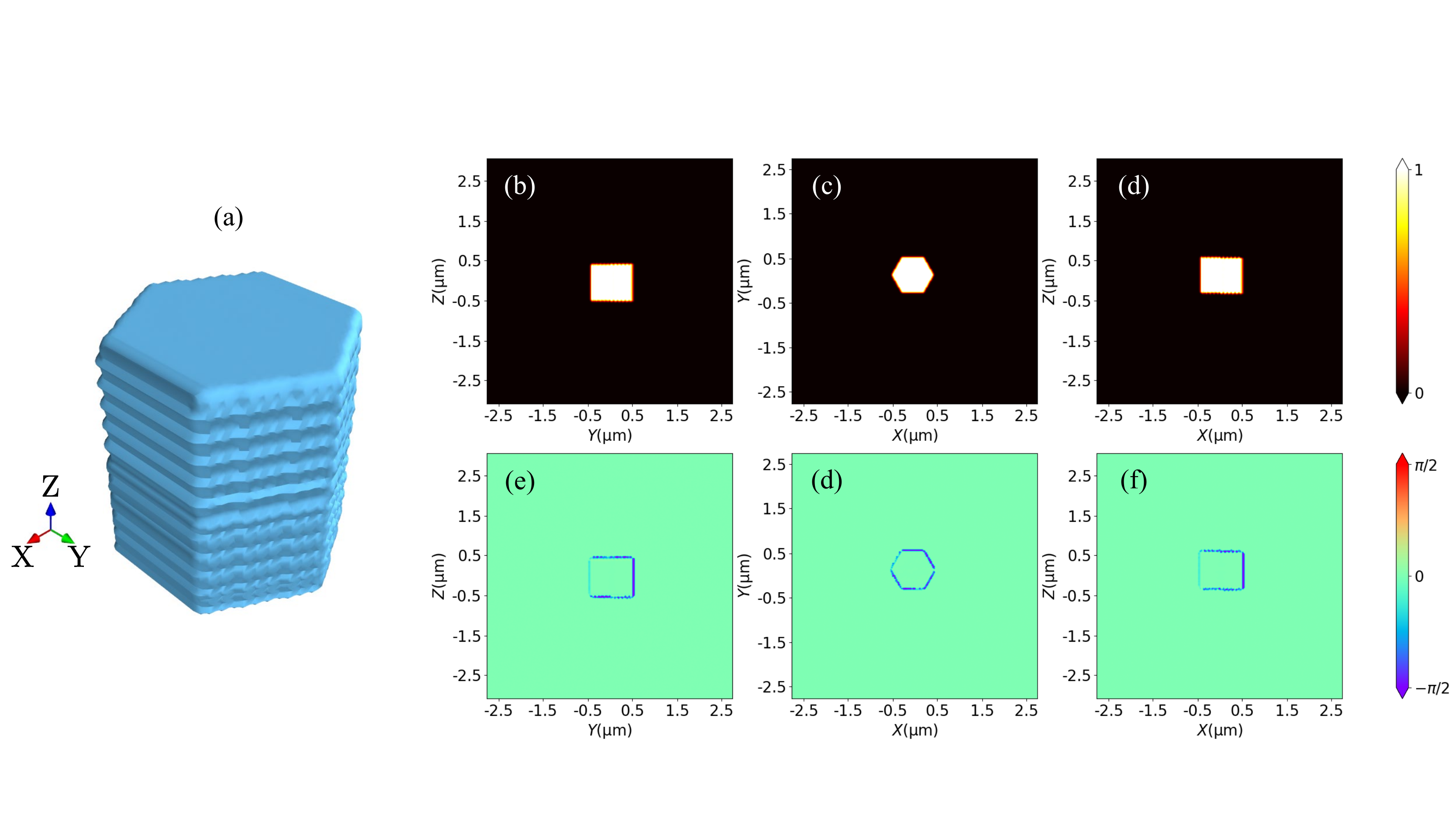}
	\caption{\label{fig4}The reconstruction results. (a) The reconstructed 3D sample density. (b)-(d) The central slices of the reconstructed 3D sample density along three planes of real space, (a)-(c) are the Y-Z, X-Y, X-Z planes, respectively. (e)-(g) The central slices of the reconstructed 3D sample phase along three planes of real space, (d)-(f) are the Y-Z, X-Y, X-Z planes, respectively.}
\end{figure*}

In this case, we simulate the bulk polycrystalline Si material of size $2\ \mathrm{\upmu m}\times 2\ \mathrm{\upmu m}\times 4\ \mathrm{\upmu m}$ (the coherent beam is incident in the direction of $4\ \mathrm{\upmu m}$) as a set of grid points in space, where the material electron density is set to 2 (the phase is set to 0 indicates that the sample is strain-free) and the density of the outer points is set to 0. As shown in Fig. \hyperref[{fig3}]{3(a)}, in the direction of $\mathbf{k}_{\mathbf{i}}$, there is a regular hexagonal crack region with a size of $1\ \mathrm{\upmu m}\times 1\ \mathrm{\upmu m}\times 1\ \mathrm{\upmu m}$ in the center of the material (at the depth of 2 $\mathrm{\upmu m}$), and a cuboid crack region with a size of $0.4\ \mathrm{\upmu m}\times 0.4\ \mathrm{\upmu m}\times 1.4\ \mathrm{\upmu m}$ in different orientations at both ends (at the depth of 0.5 $\mathrm{\upmu m}$ and 3.5 $\mathrm{\upmu m}$, respectively). The density of these three crack regions is set to 1. The material is illuminated with x-rays of energy of 13 keV ($\lambda=0.95\rm \mathring{A}^{-1}$), and a small angular rocking curve collected in the vicinity of the (111) Bragg peak consisting of 256 even intervals over $\sim \pm 0.5^{\circ}$ (the angle step size $\sim 0.004^{\circ}$) results in a series of effectively parallel 2D slices from which a 3D reciprocal space map of the Bragg peak and its surrounding fringes can be resolved so as to satisfy the Nyquist sampling criterion. And we simulate a square detector, which is oriented in such a way that the center of the detector makes an angle of $ 2\theta_B= 17.5^{\circ}$ with the incoming x-ray beam, with $256\times 256$ pixels of size 80 $\mathrm{\upmu m}$ of 5.64 m away from the sample to ensure that the diffraction pattern was oversampled. A Be-based CRL is positioned 10 cm from the sample and contain 25 2D Be lens with an apex radius of curvature $R = 40\ \mathrm{\upmu m}$ and center-to-center distance between successive lenslets $T = 16\ \rm{mm}$, yielding an focal length of CRL $f_N=38\ \mathrm{cm}$, an effective aperture $Y=253\ \mathrm{\upmu m}$, DoF $\Delta \sim 1\ \mathrm{\upmu m}$, and FoV $\mathrm{\Lambda_o} \sim 1\ \mathrm{\upmu m} \times 1\ \mathrm{\upmu m}$. With these configurations, we simulate the separate imaging of the regular hexagonal crack region at the center of the bulk material. In our simulation, the Fresnel number $Fr=0.0019\ll 1$ indicates that the diffraction pattern is measured in the far field, and the diffraction pattern is oversampled by a factor of approximately 5 in all directions. By stacking a series of 2D slices near the (111) Bragg peak obtained along the rocking curve, the 3D diffraction intensity is shown in Fig. \hyperref[{fig3}]{3(b)}. Any non-zero displacement field of the atoms in the crystal from their ideal lattice positions (strained crystal) will lead to asymmetric intensity distribution at the Bragg spots \cite{https://doi.org/10.1002/adma.201304511}. The three central slices of the simulated 3D intensity are shown in Fig. \hyperref[{fig3}]{3(c)}. The clear symmetry of the diffraction intensity indicates that the crystal is strain-free.

The 3D rendering of the reconstructed sample density with the resolution of $\sim 20\ \mathrm{nm}$ is shown in Fig. \hyperref[{fig4}]{4(a)}. The central slices of the reconstructed 3D density along three planes of real space (Y-Z, X-Y and X-Z planes) are shown in Figs. \hyperref[{fig4}]{4(a)}-\hyperref[{fig4}]{4(c)} respectively, and the central slices of the reconstructed phase along three planes of real space (Y-Z, X-Y, X-Z planes) are shown in Figs. \hyperref[{fig4}]{4(d)}-\hyperref[{fig4}]{4(f)} respectively. It can be seen that the density and phase within the reconstructed sample are uniform, consistent with the initial set of samples. This verifies that the introduction of CRL corresponds to the modification of the far-field diffraction phase of the object, that is, the influence of aberration on the diffraction pattern and reconstruction quality can be ignored. During the transformation between orthogonal space and measurement space, the spatial grid is not fine enough, resulting in regular stripes along the $\mathbf{k}_{\mathbf{i}}$ direction on the simulated sample surface. The 3D rendering of the reconstruction shows that the stripe was preserved during the reconstruction. It can be known from the slice diagram that the characteristic width of this stripe is about 20-40 nm. In addition, the limit of the NA of CRL on the spatial resolution has been given by Eq. \eqref{resolution}. In this case, the limit of spatial resolution is $r\ =\ 25\ \mathrm{nm}$, which is consistent with the resolution of our reconstruction results. Using higher energy x-rays or optimizing the lens configuration is expected to improve the resolution, but it also leads to the need to make compromises in parameters such as resolution, DoF or FoV.

It should be noted that in some experiments such as gas-gun loading method limits the experimental sample scale to a lower limit of 100 $\mathrm{\upmu m}$, usually which is upper limit of micro-scale for most of solid materials. Therefore, the scheme demonstrated in this paper needs to be extended to the application of mesoscale samples. This is not a simple and straightforward extension when imaging submicron-sized structures within the tens to hundreds of $\mathrm{\upmu m}$-sized bulk materials. Firstly, thicker samples require higher energy and better coherent light sources, such as XFEL, and secondly, the effect of scattered light from non-imaging regions on scattered light from imaging regions needs to be taken into account. Moreover, 3D dynamic imaging of bulk materials is particularly important to understand the evolution processes of microscopic scale structures, but it is not compatible with sample rotation methods. The XMPI scheme proposed by Villanueva-perez \emph{et al.} is a promising 3D single-shot method. XMPI has demanding requirements on the brilliance of the light source and is thus envisioned as an imaging method for XFELs. In addition, the peak brightness of XFEL is nearly ten orders of magnitude beyond conventional synchrotron sources, and its pulse duration can reach 10 fs, enabling imaging of the structure of matter at the atomic size and timescales \cite{Emma2010, Ishikawa2012}. The combination of the method demonstrated in this paper with XMPI requires the addition of optical lens to each of the multiple beam paths generated by the beam splitter through Laue diffraction. This complex optical system is still under development.

\section{Conclusion}
In conclusion, we study the effects of DoF and FoV of the optical lens in objective BCDI, where DoF and FoV together limit the diffraction volume and DoF is critical for limiting the thickness of the imaging region, so as to achieve the separate high resolution imaging of selected crystals in bulk polycrystalline materials. Separate imaging of submicron-sized crystal in several $\mathrm{\upmu m}$-sized bulk Si material is verified by numerical simulations and showed considerable reconstruction quality. Based on XFEL, this scheme can be extended to the 3D dynamic imaging of micron scale imaging of bulk materials with tens to hundreds of $\mathrm{\upmu m}$ size. It is important to understand the evolution of the microscopic scale structure of materials during dynamic compression.

\bibliographystyle{apsrev4-2.bst} 
\bibliography{ref}

\end{document}